# Methodological Foundation of a Numerical Taxonomy of Urban Form

Martin FLEISCHMANN

Department of Geography and Planning, University of Liverpool, Roxby Building, Liverpool, L69 7ZT, United Kingdom. m.fleischmann@liverpool.ac.uk; +44(0)7482 082 1247; Corresponding author

Department of Architecture, University of Strathclyde

Alessandra FELICIOTTI

Department of Architecture, University of Strathclyde, a.feliciotti@strath.ac.uk

Ombretta ROMICE

Department of Architecture, University of Strathclyde, ombretta.r.romice@strath.ac.uk

Sergio PORTA

Department of Architecture, University of Strathclyde, sergio.porta@strath.ac.uk

**Acknowledgements**

This work was supported by the Axel and Margaret Ax:son Johnson Foundation as part of "The Urban Form Resilience Project" at the University of Strathclyde. Proprietary data used for the validation were provided by the Institute of Planning and Development Prague.

## Abstract

Cities are complex products of human culture, characterised by a startling diversity of visible traits. Their form is constantly evolving, reflecting changing human needs and local contingencies, manifested in space by many urban patterns.

Urban Morphology laid the foundation for understanding many such patterns, largely relying on qualitative research methods to extract distinct spatial identities of urban areas. However, the manual, labour-intensive and subjective nature of such approaches represents an impediment to the development of a scalable, replicable and data-driven urban form characterisation.  Recently, advances in Geographic Data Science and the availability of digital mapping products, open the opportunity to overcome such limitations. And yet, our current capacity to systematically capture the heterogeneity of spatial patterns remains limited in terms of spatial parameters included in the analysis and hardly scalable due to the highly labour-intensive nature of the task. In this paper, we present a method for numerical taxonomy of urban form derived from biological systematics, which allows the rigorous detection and classification of urban types. Initially, we produce a rich numerical characterisation of urban space from minimal data input, minimizing limitations due to inconsistent data quality and availability. These are street network, building footprint, and morphological tessellation, a spatial unit derivative of Voronoi tessellation, obtained from building footprints. Hence, we derive homogeneous urban tissue types and, by determining overall morphological similarity between them, generate a hierarchical classification of urban form. After framing and presenting the method, we test it on two cities - Prague and Amsterdam - and discuss potential applications and further developments. The proposed classification method represents a step towards the development of an extensive, scalable numerical taxonomy of urban form and opens the way to more rigorous comparative morphological studies and explorations into the relationship between urban space and phenomena as diverse as environmental performance, health and place attractiveness.

*Keywords*: urban morphometrics, classification, numerical taxonomy, urban morphology

## Introduction

Cities' visual diversity is astounding. Indeed, when comparing their spatial form, marked differences can be clearly observed at all scales. And yet, despite these variations, their heterogeneous fabrics share geometric characteristics, which make it possible to compare them to one another through the analysis of their constituent elements and, to recognise patchworks of distinct *urban tissues* within each city.

The endeavour of capturing these multifaceted spatial patterns has been the object of investigation across multiple disciplines. Notably, building on research in geography (Conzen, 1960) and architecture (Muratori, 1959), the discipline of urban morphology devote over 60 years to explore recurrent patterns within urban forms in cities all over the world, aiming at their definition, classification and characterisation (Kropf 1993, 2014; Oliveira 2016).

Further research has focused on classification of morphological elements into "types". This includes the series of works by Steadman (Steadman, Bruhns and Holtier, 2000; Steadman, Evans and Batty, 2009) on the classification of buildings based on a handful of empirically measured geometrical parameters as well as the work by Marshall (2005) on the classification of street pattern types.

And whilst these contributions are heterogeneous both in terms of object of interest (i.e. building, street, urban tissue), method (i.e. qualitative vs quantitative) and aim of the classification (i.e. energy performance, historical origin, design paradigm), they mark important attempts at classifying the variations of individual elements – buildings (Steadman et al. 2000, Steadman et al. 2009) – or aggregations of individual elements - street patterns (Marshall, 2005) – making up the of spatial form of cities through geometrical analysis. As such they mark steps towards a more rigorous study of relationships between different urban configurations.

Yet, our current capacity to systematically capture the heterogeneity of spatial patterns remains limited. Most existing research in urban morphology relies on highly-supervised, expert-driven and labour-intensive qualitative methods both in the data preparation process and in the design of the analysis. As a result, most existing works are hardly scalable due to the considerable amount of manual work required to prepare the input data and tend to focus on the analysis of relatively few spatial parameters.

Recently however, advances in geographic data science, combined with growing availability of geospatial data, triggered a data-driven stream of urban morphology studies, named *"urban morphometrics"* (e.g. Gil et al. 2012, Dibble et al. 2019, Araldi & Fusco 2019, Bobkova 2019). Within this line of research, the paper aims to address the need for more systematic, scalable and efficient method for the detection and classification of morphological patterns. To this end, after presenting a brief literature review on urban form classification and specifying the requirements for a rigorous classification method, we

- present an original quantitative methodology for the systematic unsupervised classification of urban form patterns and ground it on the theory of phenetics and numerical taxonomy in biological systematics.
- apply the proposed methodology to two exploratory case studies, as proofs of concept aimed at providing an illustration of the method and some of its potential theoretical impacts and technical shortcomings.

More specifically, we will first frame the proposed approach to urban form classification within numerical taxonomy, which seeks to describe and classify species and taxa based on morphological similarity (Sneath & Sokal, 1973). To build this methodological parallel between the (a-biotic) system of urban form and biology, we a) re-frame the constituent elements of urban forms as the building blocks of the method, 2) describe how to identify structurally homogeneous urban form types (or "taxa") and 3) measure their hierarchical relationship based on phenetic similarity, delivering a systematic numerical taxonomy of urban form. Finally, we test the proposed method on two major European cities characterised by various types of urban fabric originating from different historical stages: Prague, CZ and Amsterdam, NL.

We conclude discussing validation findings, highlighting potential theoretical impact of the proposed method and discussing methodological limitations.

## Existing models of urban form classification

The primary aim of classification is to reduce the complexity of the world around us. Many urban form classification methods exist at building (Steadman et al, 2000, Steadman et al. 2009, Schirmer & Axhausen, 2015), street (Marshall, 2005) neighbourhood (Soman et al., 2020) and city (Louf & Barthelemy, 2014) scales, varying conceptually and analytically both in terms of focus scale - e.g. global, (Angel et al. 2012) vs local (Guyot et al. 2021), analytical approach – e.g. quantitative vs. qualitative, and aim of the classification. Structurally, the simplest forms involve *flat* classifications, where the relationship between types is unknown. These are either *binary* like *organized* vs. *unorganized* neighbourhoods (Dogrusoz & Aksoy, 2007), or *multi-class,* as Caruso et al.'s (2017) 4-class clustering based on inter-building distance, or Song and Knapp's (2007) 6-class neighbourhood typology based on factor analysis and K-means of 21 spatial descriptors, or the "multiscale typology" by Schirmer & Axhausen, (2015) identifying four flat classes based on centrality and accessibility. More complex classifications involve *hierarchical* methods (*taxonomies*), which organise classes based on their mutual relationships like Serra et al. (2018)'s hierarchical taxonomy of *neighbourhoods* built according to 12 morphological characters of street network, blocks and buildings, and the work by Dibble et al. (2019) who hierarchically classify portions of urban area enclosed by main streets. More granular approaches include the work by Araldi & Fusco (2019), who classify *street segments* using 21 morphometric characters derived from street networks, building footprints and digital terrain model and research by SMOG at Chalmers University (Berghauser Pont et al., 2019a; Berghauser Pont et al., 2019b; Bobkova et al., 2019) that classifies morphological elements of plots, streets and buildings through a handful of morphometric characters.

Other approaches employ morphometric assessment to predict pre-defined typologies of buildings, streets or larger areas (Marshall, 2005, Hartmann et al., 2016; Neidhart and Sester, 2004; Steiniger et al., 2008; Wurm et al., 2016). These validate morphometrics in classification of urban form, even though the typology itself is defined differently. Related to this are Urban Structural Type classifications reviewed by Lehner & Blaschke (2019), and detection of Local Climate Zones (Stewart & Oke, 2012; Taubenböck et al., 2020).

Whilst the list does not aim to be exhaustive of all contributions it nevertheless provides an overview of the state of the art in urban form classification research. Specifically, it highlights how each of these method shows shortcomings in scalability (the ability to analyse large areas while retaining the detail), transferability (the ability to apply to different contexts), robustness (the ability to remain unaffected by small imprecision of the input data or measurement), and extensiveness (i.e. the bias induced by a small number of variables), or interpretative flexibility (i.e., missing relations between classes). This leaves a methodological gap in morphometric classification of built environment hindering the development of universal taxonomy of urban form.

## Method: Building a taxonomy of urban form

The problem of classification of urban patterns based on geometrical resemblance is not dissimilar, conceptually speaking, to the work of early biologists seeking to classify biotic species and taxa based on morphological similarity. This was indeed the primary aim of numerical taxonomy (and generally *phenetics*), established in biology in the second half of the 20th century (Sneath & Sokal, 1973).

Whilst DNA sequencing and *phylogenetics* have now largely replaced morphometrics in modern biological taxonomy, we can take advantage of the latter for the study of urban form. Very much like the study of organismal phenotypes and the statistical description of biological forms were instrumental to the separation of individuals (and species) into recognisable, homogeneous groups (Raup,1966), extending numerical taxonomy to the study of urban form offers an operationally viable and reliable conceptual and methodological framework for a systematic classification of homogeneous urban form types.

And yet, whilst this possibility has always fascinated urban scholars in an analogic sense (Philip and Steaman, 1979), a rigorous methodological parallel between numerical taxonomy and urban form classification is a matter of pioneering research.

One of the first authors to explicitly use numerical taxonomy on urban form was Dibble et al. (2019) who, notwithstanding operational limitations, measured a large number of geometrical parameters of fundamental morphological elements (buildings, streets, plots etc) to test the applicability of the approach in urban morphology. However, their method requires predefined boundaries of urban types, is extremely data demanding and is not possible to do without manual measuring. Despite that, it paved the conceptual way for further research including the one presented in this paper.

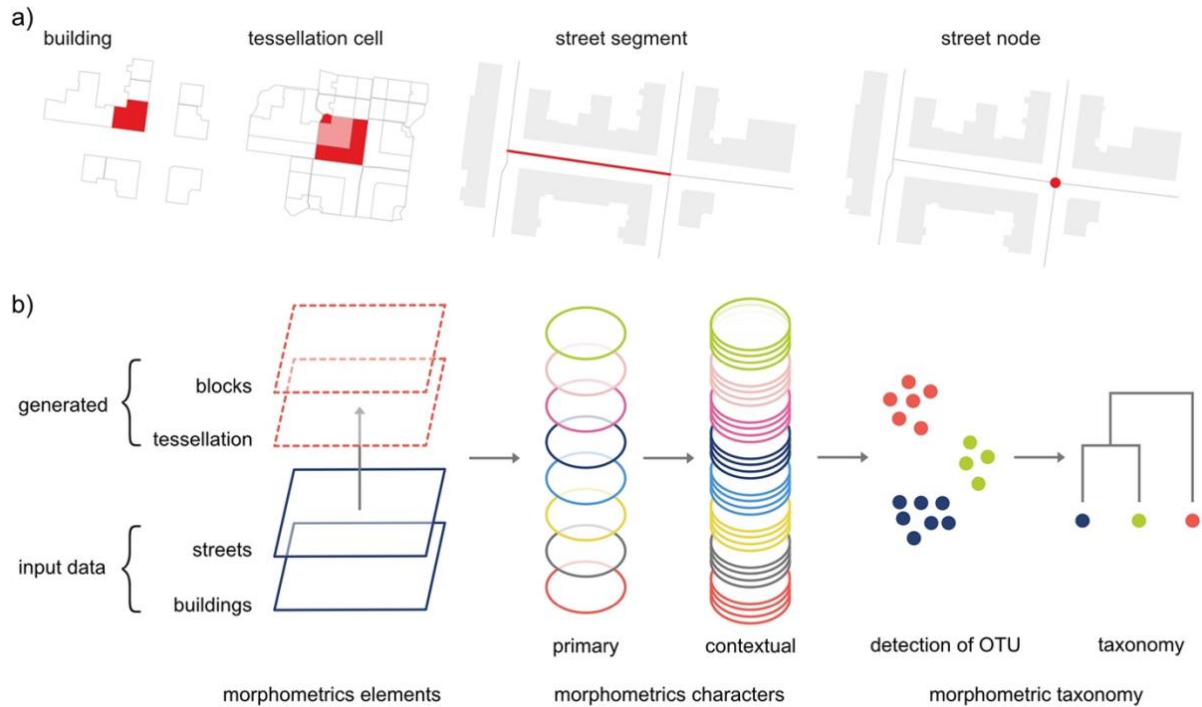

**Figure 1: a)** Fundamental morphometric elements: *building footprint*, *tessellation cell* (derived from building footprints) and *street* (segment and node from centrelines). **b)** Diagram illustrating the workflow of the proposed method. From input data (buildings, streets) are derived generated elements (tessellation, blocks). All elements are used to measure primary morphometric characters. Each of them is then represented as 4 contextual characters that are used as an input of the cluster analysis. Finally, resulting classes are organised in a taxonomy.

Before we can define the method for numerical taxonomy of urban form we need to establish

the building blocks, namely *structural morphometric elements*, or the urban form counterpart

of the individual and its body in biology (Sneath & Sokal, 1973), and *operational taxonomic*

*unit* (OTU), or else the unit forming the lowest ranking taxa, which in biology is individuals or populations depending on taxonomic level.

**Morphometric elements**

Urban morphologists generally agree on three fundamental elements: buildings, plots and streets (Kropf, 2017; Moudon, 1997). To make our method scalable it is imperative that, when these are translated into operational and measurable morphometric elements, i.e., vector features in GIS data, they maintain their meaning with minimal data input, hence maximising data accessibility and consistency.

From a morphometric standpoint, this is relatively straightforward for *streets* and *buildings* due to their conceptual simplicity: buildings can be represented as *building footprint* polygons (with the attribute of building height) at Level of Detail 1 (Biljecki et al., 2016) whilst streets as *network* centrelines, cleared of transport planning-related structures. The same is more complicated for the *plot,* particularly at large scale, due to its highly polysemic nature (Kropf, 2018) and ambiguous structuring role in contemporary urban fabrics (Levy, 1999).

To avoid the plot's inconsistencies, we use *morphological tessellation*, a polygon-based derivative of Voronoi tessellation obtained from building footprints proposed by Fleischmann et al. (2020b) after Hamaina et al. (2012) and Usui & Asami (2013) and the *morphological cell*, its smallest spatial unit which delineates the portion of land around each building that is closer to it than to any other but no further than 100m. As such, the morphological tessellation captures the topological relations between individual cells and influence that each building exerts on the surrounding space (Hamaina et al., 2012), regardless of historical origin, thanks to its contiguity throughout the analysis space (**figures 1a and 2**). Furthermore, being generated solely from building footprints, it does not increase data reliance. However, as such, it does not have the ability to represent unbuilt areas and empty plots and does not serve as a substitute for plot in general terms as it does not have the same structural role. Morphological tessellation is a purely analytical element.

### *Operational taxonomic unit*
In biology the *operational taxonomic unit* (OTU) is intuitive (individual organism). The same is, however, not true for urban form. In urban morphology, this can be associated to the concept of "morphological regions" (Oliveira & Yaygin, 2020), "urban tissues" (Caniggia & Maffei, 2001; Kropf, 1996) or "urban structural types" (Lehner & Blaschke, 2019; Osmond, 2010), or else *"a distinct area of a settlement in all three dimensions, characterized by a unique combination of streets, blocks/plot series, plots, buildings, structures and materials and usually the result of a distinct process of formation at a particular time or period"* (Kropf 2017, p.89).

From a morphometric standpoint, adopting the concept of "urban tissue" as the OTU has two main advantages. First, being grounded on the notion of *homogeneity,* its definition can be configured as a typical problem of cluster analysis: homogeneous urban tissues are hence derived from the analysis of recurrent similarities/differences in the morphometric characters of their constituent urban elements. Furthermore, as size and geometry of each urban tissue are determined by internal homogeneity rather than pre-defined boundaries, the Modifiable Aerial Unit Problem is minimised (Openshaw, 1984).

Having the elements defined, the method proposed here can be split into five consecutive steps illustrated on **figure 1b**: 1) generation of morphological elements, 2) measurement of primary morphometric characters, 3) measurement of contextual character, 4) detection of OTUs, 5) taxonomy. Step 1 is covered above and the remaining steps are outlined in the following sections.

**Morphometric characters**
*Morphometric characters* are the measurable traits of each morphometric elements - the "wing's length" or "beak's dimension" in biology. The definition of measurable *morphometric characters* is key for cluster analysis and captures the cross-scale structural complexity of different urban tissues. To this end, building on an earlier literature review (Fleischmann et al., 2020a), we use six categories of morphometric characters - dimension, shape, spatial distribution, intensity, connectivity, diversity.

These characters allow a numerical description of morphometric elements (street segments, building footprints and tessellation cells) within any urban fabric, by capturing the relationships between them and their immediate surroundings. They are measured at three topological scales: *small* (element itself), *medium* (element and its immediate neighbours) and *large* – the element and its neighbours within $k$-th order of contiguity. Spatial contiguity can either be kept constrained by enclosing streets (the equivalent of an urban block) or left unconstrained (see the Supplementary Material 1 for further details).

Considered morphometric characters are of two types: *primary* and *contextual*. Primary characters measure geometric and configurational properties of morphometric elements (buildings, streets, and cells) and their relationships (at all scales). By abundantly representing all six morphometric categories this set is *extensive*. Accordingly, starting from as broad a set of unique variables identified by Fleischmann et al. (2020a), we shortlist 74 characters (**table S1** in the Supplementary Material), following rules by Sneath & Sokal (1973) to minimise potential collinearity and limit redundancy of information, while retaining the universality of the method.

Primary characters describe morphometric elements and their immediate neighbourhood rather than their *spatial patterns*. As such, when employed for cluster analysis they may result in spatially discontinuous classes. Urban tissues are defined by their internal homogeneity, but it can, and often is, be the homogeneity of heterogeneity. In other words, the tissue may be

defined by the combination of small and large buildings or various shapes, and we need to

capture these characteristics. Thus we derive a set of spatially lagged *contextual characters*

describing the *tendency* of each primary character in its *context*. The term "context" is here

defined as topological aggregation of morphological cells within *three topological steps* from

each given cell $C_i$, an empirically determined value large enough to capture a cohesive pattern

over a relatively wide spatial extent but small enough to generate sharp boundaries between

different patterns (**Figure 2**). The notion of "tendency" is in turn quantified through four

values:

1. *Interquartile mean (IQM)*, the most representative value cleaned of the effect of
   potential outliers.

2. *Interquartile range (IQR)*; as local measure of statistical dispersion, describes the
   range of values cleaned of outliers:

$$IQR_{ch} = Q3_{ch} - Q1_{ch},$$

where $Q3_{ch}$ and $Q1_{ch}$ are is the third and quartiles of the primary character.

3. *Interdecile Theil index (IDT)*, describes the local (in)equality of distribution of values:

$$IDT_{ch} = \sum_{i=1}^{n} \quad (\frac{ch_i}{\sum_{i=1}^{n} ch_i} ln \, [N \frac{ch_i}{\sum_{i=1}^{n} ch_i}]),$$

where $ch$ is the primary character.

4.  *Simpson's diversity index (SDI)*, captures the local presence of classes of values compared to the global structure of the distribution:

$$SDI_{ch} = \frac{\sum_{i=1}^{R} n_i(n_i-1)}{N(N-1)},$$

where $R$ is richness, expressed as number of bins, $n_i$ is the number of features within $i$-th bin and $N$ is the total number of features.

Of these, the first captures the *local central tendency* and the latter three the *distribution of values* within the third order of contiguity from each cell.

Each primary character is used as an input for each contextual option. The full set of morphometric characters hence includes 74 primary plus 296 contextual characters (74x4), totalling 370 characters. These are computed using the bespoke open-source Python toolkit momepy (Fleischmann 2019), ensuring the full replicability and reproducibility of the method.

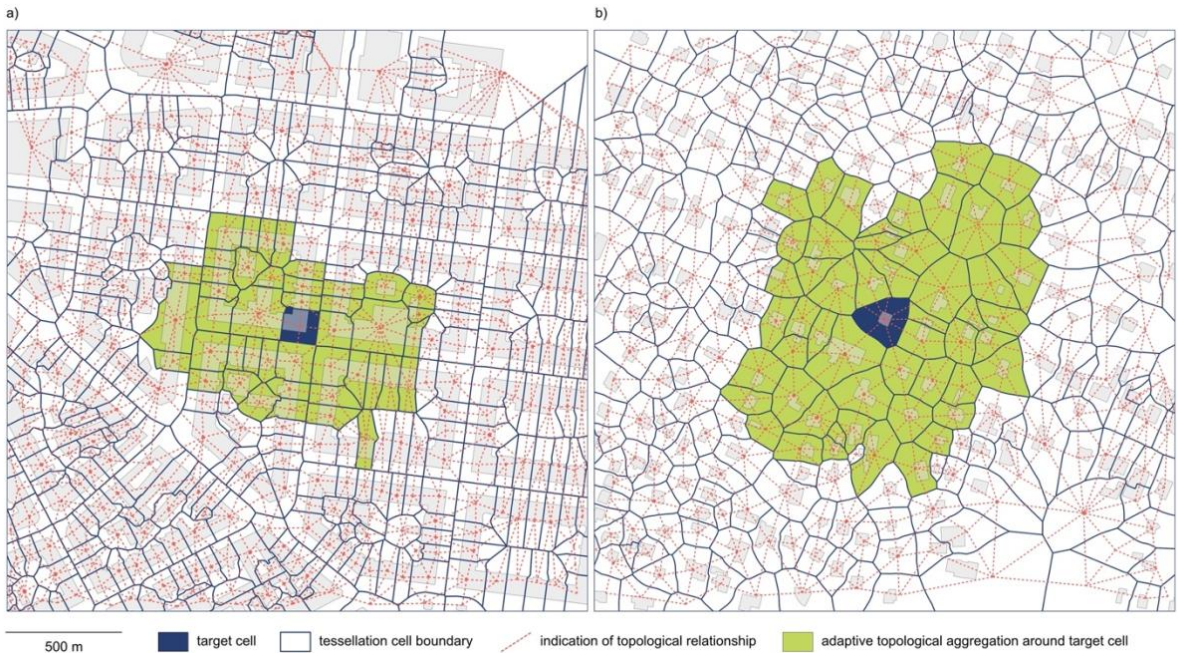

**Figure 2:** Morphological tessellation's adaptive topological aggregation; "context" is defined as all cells within third order of contiguity in Prague: a) compact perimeter blocks, b) single family housing.

## Detection of the operational taxonomic unit

Only contextual characters' values are input to cluster analysis that identifies *urban form types*. Identifying OTUs as clusters of fundamental entities closely mirrors a *mixture problem* in biology, which identifies populations within samples and classifies at population level (Sneath & Sokal, 1973). Since contextual characters are spatially lagged, they are spatially autocorrelated by design, thus avoiding computationally expensive spatial constraint models (Duque et al., 2012). We mitigate potential over-smoothing of the boundaries by basing contextual characters on truncated values (with the exception of SDI), which eliminate outliers' effect and define boundaries more precisely.

The most suited clustering algorithm is Gaussian Mixture Model (GMM), a probabilistic derivative of *k*-means (Reynolds, 2009) tested in a similar context by Jochem et al. (2020). Unlike the *k*-means itself, GMM does not rely only on squared Euclidean distances and is more sensitive to clusters of different sizes. GMM assumes that a Gaussian distribution represents each dimension of each cluster. Hence the cluster itself is defined by a mixture of Gaussians. The output of GMM are cluster labels assigned to individual tessellation cells.

The ideal outcome of cluster detection would equate clusters to distinct taxa of urban tissues. Because the definition of urban tissue (Kropf, 2017) does not specify the threshold beyond which two similar parts of a city cluster in the same tissue, it is difficult to equate clusters to taxa. We resolve this by estimating the number of clusters, required by GMM clustering method, on the goodness of fit of the model, measured using Bayesian Information Criterion (BIC) (Schwarz & others, 1978) based on the "elbow" of the curve.

**The foundation of taxonomy**
To classify urban form types, we use *Ward's minimum variance hierarchical clustering* previously applied in urban morphology (Dibble et al., 2019; Serra et al., 2018). Here, each urban form type is represented by its centroid (mean of each character across cells with the same label); Ward's algorithm links observations reducing increase in total within-cluster variance (Ward Jr, 1963). The classification is represented through a dendrogram capturing the *cophenetic relationship* between observations (i.e., morphometric similarity), forming the foundation of our taxonomy.

**Validation theory**

For validation, we study our taxonomy in relation to other urban dynamics with which some form of relation is expected. In urban morphology theory and *qualitative* evidence suggests that different urban patterns emerge in areas of different *historical origins* or else belonging to different *"morphological periods"* (Whitehand et al., 2014). This notion has also been observed *quantitatively* in the urban fabric (Boeing, 2020; Dibble et al., 2019; Porta et al., 2014, Fleischmann et al., 2021) as well as in land use patterns (Castro et al., 2019) of cities and is inherently embedded in our OTU.

We validate our classification against three datasets: 1) historical origins; 2) predominant land-use patterns, and 3) qualitative classification of urban form adopted in official planning documents. We use the same method, based on cross-tabulation, resulting in statistical analysis using chi-squared statistics and related Cramér's *V* (Agresti, 2018). The model is considered valid if a significant relationship is found between proposed classification and three additional datasets and if similar performance is shown across different case studies.

**Case study**

We test the proposed method in two historical European cities: Prague, CZ and Amsterdam, NL. Prague's analysis area is defined by its administrative boundary, which extends beyond its continuous built-up area to minimise the "edge-effect" of the street network (Gil, 2016). Amsterdam's analysis area is defined by its contiguous urban fabric, extending beyond the city's administrative boundary. The morphological data (buildings, streets) for Prague case study were obtained from city's open data portal (https://www.geoportalpraha.cz/en), while the validation layers were provided by Prague Institute of Planning and Development. The morphological data for Amsterdam are obtained from 3D BAG repository (Dukai, 2020) and Basisregistratie Grootschalige Topografie(http://data.nlextract.nl/)

## Results: Taxonomy of Prague and Amsterdam

We measure all *74 primary characters* in both Prague and Amsterdam, associated to each morphological cell, and subsequently generate *296 contextual characters* as input to cluster analysis.

**Cluster analysis in Prague**

Based on BIC results (**figure S5** in the Supplementary Material), GMM clustering identifies 10 clusters (**figure 3a**). At a visual inspection, clusters appear well defined and able to reflect homogenous forms, their contiguity resulting from contextual characters' patterned nature.

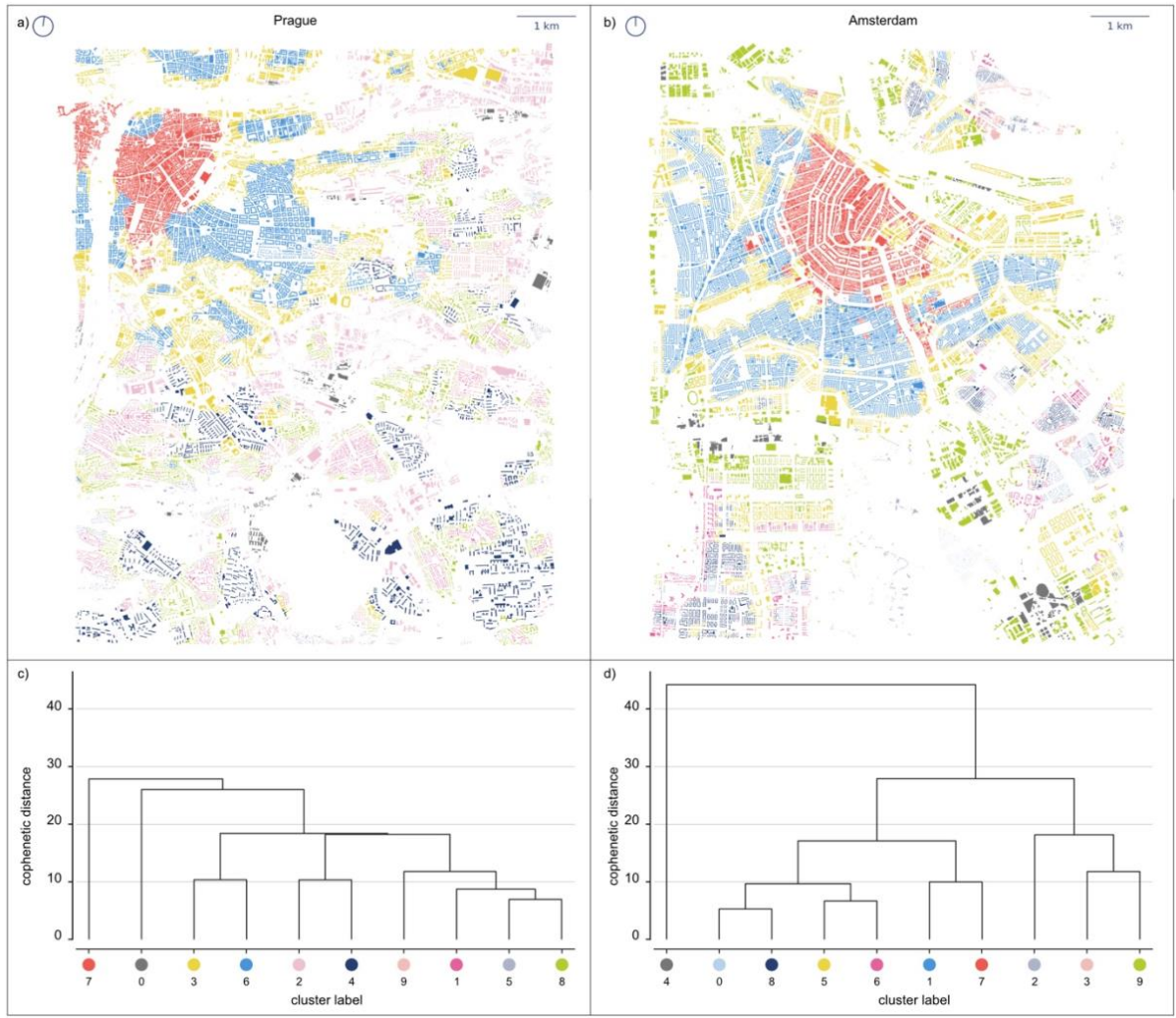

**Figure 3:** Spatial distribution of detected clusters in central Prague (a) and central Amsterdam (b) accompanied by dendrograms representing the results of Ward's hierarchical clustering of urban form types in Prague (c) and Amsterdam (d). The y-axis shows the cophenetic distance between individual clusters, i.e., their morphometric dis-similarity. The full extent of case studies is shown in figures S7 and S8 in the Supplementary Material.

Starting from the historical core of Prague (top left), we first identify the medieval urban form (7), then the compact perimeter blocks of Vinohrady neighbourhood (6,), and the fringe areas (3). Towards South and East, we note low-rise tissues (8, 1) and modernist developments (4).

Drawing purely from visual observation and personal knowledge of the city of Prague, identified clusters appear to nicely capture meaningful urban form types.

**Cluster analysis in Amsterdam**
In Amsterdam, BIC indicates the optimal number being 10 clusters, similarly to Prague.

As in Prague, the geography of clusters shows seemingly meaningful results (**figure 3b**). For example, cluster 7 captures the city's historical core up to the Singelgracht canal. The cluster 1 reflects well-known shifts in planning paradigms with the rise of New Amsterdam School (Panerai et al., 2004) forming the early 20[th] century south expansion. Once again, under preliminary observation, identified clusters capture meaningful spatial patterns.

**Numerical taxonomy**
The centroid values of each cluster, obtained as the mean value of each contextual character, are used as taxonomic characters in Ward's hierarchical clustering. Resulting relationship between centroids represents the relationship between clusters (**figure 3c**).  The dendrogram's horizontal axis represents detected clusters, while the vertical axis their *cophenetic distance* (i.e., morphological dissimilarity ): the lower the connecting link of two clusters, the higher their similarity.

Prague's dendrogram contains 10 clusters, illustrating the uniqueness of the spatial pattern of the medieval city (7), forming the first bifurcation and independent branch. The similar situation is with the cluster covering industrial areas (0) being dissimilar to other clusters. Further in the dendrogram, we can see branches with regular perimeter blocks (6) and their fringe areas (3), unorganised development of modern era (4, 2) or a branch featuring residential areas of low density (9, 1, 5, 8).

The dendrogram of Amsterdam urban form (**figure 3d**) shows similar characteristics, with bifurcations distinguishing nested levels of spatial variations.

In the classification maps shown in figure 3,types are colour-coded to highlight distinctions at individual cluster's level. However, we can instead colour-code according to clusters' similarity. Because the dendrogram shows several major bifurcations at different levels of cophenetic distance indicating distinct higher-order groups of clusters, by colouring each cluster *in the map* according to the branch it belongs to *in the dendrogram* and using different hues to distinguish between lower-level clusters in each branch, we distinguish hierarchies based on cophenetic distance.

We can further combine the two cities' clusters in one shared dendrogram (**figure 4c**).  Urban form types from both pools appear regularly distributed in the lowest orders of the tree, showing a similar spatial structure emerging in both cases. Remarkably, we can see the major bifurcation setting apart industrial urban forms in the combined taxonomy.

A lower order bifurcation within the main branch distinguishes between dense/compact urban form and the rest. Further lower-level subdivisions are also visible. Compared to individual ones, the combined tree shows some differences in branching: a few clusters are reshuffled and the branches themselves are slightly reorganised. This is likely to happen as more and more cities are analysed until the unified taxonomy reaches a "plateau" when enough cases are included, ultimately producing a 'general taxonomy of urban form'.

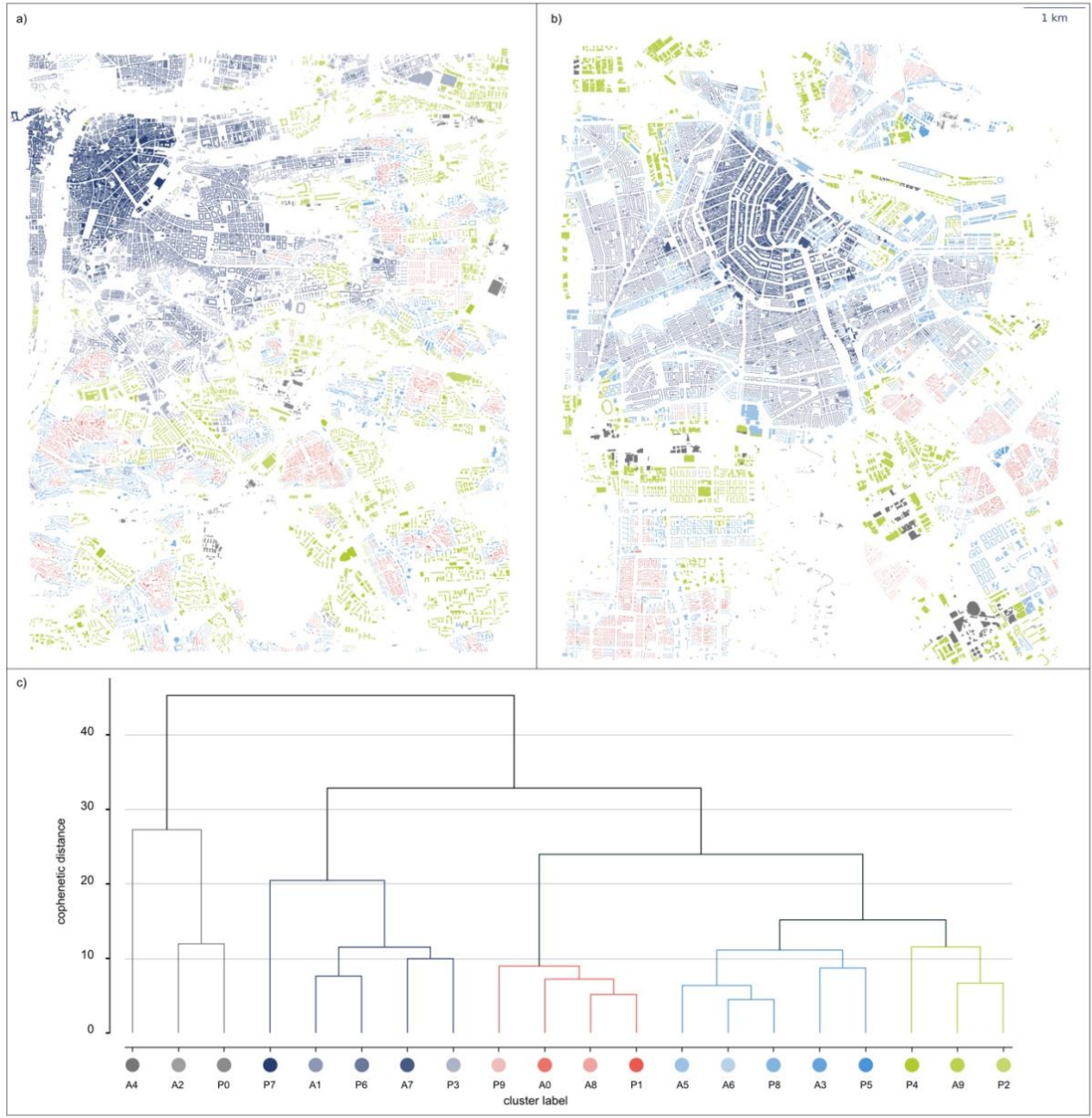

**Figure 4:** Spatial distribution of different branches of the combined dendrogram in central Prague (a) and central Amsterdam (b) accompanied by the dendrogram representing the results of Ward's hierarchical

clustering of urban form types from a combined pool of Prague and Amsterdam (c). The y-axis shows cophenetic distance between individual clusters, i.e. their morphometric dis-similarity. Branches are interpretatively coloured - the colours are then used on maps illustrating spatial distribution of these branches. The full extent of case studies is shown in figures S9 and S10 in the Supplementary Material.

The geography of Prague and Amsterdam combined taxonomy (**figure 4a, 4b**) allows cross-comparing urban form patterns by similarity (represented by similar colours). Same can be extended across a multitude of cities and regions.

**Validation**
We validate the output of numerical taxonomy against three datasets: 1) historical origins; 2) land-use patterns, and 3) qualitative classifications. All these are assessed by contingency table-based chi-squared statistics and Cramér's *V*.

In Prague, data on historical origin classifies urban areas into 7 periods: *1840, 1880, 1920, 1950, 1970, 1990, 2012*, while there are 123 categories of land use at individual building/plot level, where only 15 contain more than 1,000 buildings. We redefined prevailing land uses within the 3 topological steps of morphological tessellation: only 5 categories (*Multi-family housing, Single-family housing, Villas, Industry small, Industry large*) contain more than 1% of the dataset. We use these five and denote the rest as *Other*.

Qualitative classification is drawn from a municipal typology of neighbourhoods developed by the city for planning purposes. Each neighbourhood has specified boundaries based on its morphology and other aspects, from historical origin to social perception and qualitatively classified according to 10 types. We exclude 3 types, *hybrid* and *heterogenous*, which are non-morphological and *linear* which captures railway structures only.

Differently from Prague, the Amsterdam dataset of historical origin (Dukai, 2020) indicates each building's year of construction, starting with 1800, rather than the period of first settlement. To ensure data compatibility with the method and avoid issues with pre-1800 periods, origin dates are binned into 11 groups following Spaan and Waag Society (2015).

The resulting chi-squared and Cramér's $V$ values are reported in table S7. Contingency tables are available as tables S3 – S6. All tests indicate moderate to high association between identified clusters and the 3 sets of validation data, supporting model's validity.

Historical origin shows moderate association in both Prague ($V$=0.331) and Amsterdam ($V$=0.311). Because of the nature of data, where period of first development is not the only driver of form and we have tissues – e.g. single-family – populating multiple historical periods, a moderate association is expected. Land use ($V$=0.468) and municipal qualitative classification ($V$=0.674), tested only in Prague, indicate moderate and high association to clusters. Again, since land use is only a partial driver of urban form, moderate association supports the proposed method's potential to capture urban reality. Furthermore, relationship between morphometric types and qualitative ones sourced from local authority is the highest among validation data, reaching $V$=0.674. This seems encouraging, since both classifications aim to capture a similar conceptualisation of the built environment.

## Discussion

The proposed method hierarchically classifies urban form types according to the similarity of their morphological traits. It is numerical, unsupervised, rich in information and scalable in spatial extent. It identifies clusters of urban form as distinct *urban form types* and, within each, contiguous *urban tissues*, reflecting that in a typical city we observe tissues belonging to the same type. The method is parsimonious in terms of input data, requiring only building footprints (and height) and street networks, to generate three morphometric elements (building units, street network, morphological tessellation) and to compute the 370 morphometric characters. Such a wealth of fine-grained information allows extensively characterising each building in the study area and its adjacency and deriving distinct urban form types hierarchically organised according to similarity.

The method allows urban form analysis both in detail and at large scale, hence overcoming a methodological gap; it is fully data-driven and does not rely on (but confirms) experts' judgement other than for interpretation of BIC score. It is structurally *hierarchical,* which ensures depth along the similarity structure of urban form types and flexibility of use, according to the desired resolution of classification. Furthermore, it is *extensive,* encompassing a broad range of morphometric descriptors between major urban form components and their context; and it is *granular*, since morphometric characters are referred to each individual building.

Finally, it is *scalable* and reproducible, in that it is designed to suit well the large scale of coverage - like cities and combinations of cities - and its source code is available open-source.

Information generated with the proposed method supports applications at three different levels. First, the set of morphometric characters can be input to studies of a relationship between urban form and socio-economic aspects of urban life, e.g. via regression analysis. This includes investigations into the link between urban form and energetic/bioclimatic performance of cities, population health, gentrification and place attractiveness. Second, flat clustering with morphometric profiles can provide aggregated information on patterns without dealing with individual characters. This makes it possible to capture the overall morphological "*identity*" of an urban tissue rather than focusing on one element at the time. Third, the taxonomy brings hierarchy into classification and, as such, it can adapt its resolution to fit any question asked. In this sense, while the results of the clusters may be well-suited for fine-grained spatial analyses, by horizontally cutting the dendrogram at a desired height, it is possible to group clusters into fewer, more generalised spatial aggregations which might be better suited for analyses at coarser resolution.

Whilst parsimonious in terms of input data, our method still relies on their availability and consistency. The building footprints layer is often of sub-optimal quality level: adjacent buildings may be represented as unified polygons, misleading the method in dense areas. Building-level information on height may not be available, reducing depth of information with potentially negative effects on the quality of resulting clusters. Consistency of data across geographies may also be an issue, particularly for large spatial extents, which may require data generated independently by multiple sources.

## Conclusions

The paper presents an original data-driven approach for the systematic unsupervised classification and characterisation of urban form patterns grounded on numerical taxonomy in biological systematics and which clusters *urban tissues* based on phenetic similarity, delivering a systematic numerical taxonomy of urban form. More specifically it measures a selection of 74 primary characters from input data (*buildings*, *streets*) and derived generated elements (*tessellation* and *blocks*), each of which is represented through 4 contextual characters (*Interquartile mean*, *Interquartile range*, *Interdecile Theil index*, *Simpson's diversity index)*. These are then used as an input of the cluster analysis, resulting in a hierarchical taxonomy. Finally, the proposed approach is validated through two exploratory case studies illustrating how the resulting clustering shows a significant relationship with validation data reflecting other urban spatial dynamics.

Urban morphometrics and the proposed classification method represent a step towards the development of a taxonomy of urban form and opens to scalable urban morphology. By overcoming existing limitations in the systematic detection and characterisation of morphological patterns, the proposed approach opens the way to the large-scale classification and characterisation of urban form patterns, potentially resulting, if applied to a substantial pool of cities, in a universal taxonomy of urban form.

At the same time, the proposed approach also provides valuable tools for more rigorous comparative studies, which are fundamental to highlight similarities and differences in urban forms of different urban settlements in different contexts, and to explore the relationship between urban space and phenomena as diverse as environmental performance, health and place attractiveness and more.

## Code and data statement

Reproducible code is available at https://github.com/martinfleis/numerical-taxonomy-paper.

Data representing building footprints of Prague are available from https://www.geoportalpraha.cz/en. Remaining data on Prague case study are available from Institute for Planning and Development Prague upon request. Data representing building footprints in Amsterdam are available from http://3dbag.bk.tudelft.nl. Data representing street network in Amsterdam are available from Basisregistratie Grootschalige Topografie, BGT

(http://data.nlextract.nl/).